# "SPIN" AND "ORBITAL" FLOWS IN A CIRCULARLY POLARIZED PARAXIAL BEAM: ORBITAL ROTATION WITHOUT ORBITAL ANGULAR MOMENTUM

A.Ya. Bekshaev

*I.I. Mechnikov National University, Dvorianska 2, 65082, Odessa, Ukraine*
*E-mail address*: bekshaev@onu.edu.ua

**Abstract**

In light beams with circular or elliptic polarization, the transverse energy flow consists of the "spin" and "orbital" parts. Both of them can induce the orbital motion of microparticles suspended within the field of a light beam, and this should be taken into account in experiments on the spin-to-orbital angular momentum conversion. The character of the spin, orbital and total transverse energy flows in circular Laguerre-Gaussian beams is studied analytically; graphical representations of the flows in the beam cross section (flow maps) are calculated and analyzed. The spin circulatory flow can be directed oppositely to the orbital one and/or to the polarization handedness. As a result, the total transverse energy circulation of a beam with homogeneous circular polarization can be of different handedness in different regions of the beam cross section, which are separated by the contours of zero transverse energy flow. Regarding the particle position within the beam cross section, it can perform orbital, spinning or combined spinning-orbital motion with variable parameters. Possible applications to optical driving of microparticles are discussed.

Rotational properties of light attract the steady and growing interest in the current literature in optics (see, e.g., reviews in Refs. [1–6]). In general, these properties are associated with the circulatory flows of energy in the plane orthogonal to the beam propagation axis and are expressed by the mechanical angular momentum (AM) of the optical field that can be transmitted to other objects, e.g. microparticles [6–10]. Regarding the nature and origination of the considered rotational properties, two sorts of AM are commonly accepted [1,21]. The spin AM is inherent in light beams with circular or elliptic polarization and owes to the field vector rotations that take place in every point of the beam cross section; the orbital AM is attributed to the "macroscopic" energy circulation caused by the beam spatial configuration (for example, the screw wavefront dislocations associated with so called optical vortices [1–5,11]). Although there exist some theoretical subtleties concerning the legality of separation of the total AM of the electromagnetic field into the spin and orbital parts in general case [1,12], the notions of spin and orbital AM are suitable and physically consistent in many practical situations.

In the last few years, a considerable attention is paid to mutual exchange of the spin and orbital AM in the AM-carrying light beams, in particular, the spin-to-orbital AM conversion induced by the beam transformations causing its strong transverse inhomogeneity [13–20]. Such transformations, for example, sharp focusing [13–17] or transmitting through small apertures [18–20], are always accompanied by essential deviations from the paraxial character of the beam propagation. Under non-paraxial conditions, the unambiguous separation of the beam angular momentum into the spin and orbital parts is impossible [1,12]; however, one still can separate the contribution associated with the beam polarization state and the contribution owing to the beam spatial inhomogeneity [15,21,22]. Namely, the energy flow density (the Poynting vector time-averaged over the oscillation period) of a monochromatic optical beam can be presented in the form

$$\mathbf{S} = \mathbf{S}_C + \mathbf{S}_O \tag{1}$$

where $\mathbf{S}_C$ and $\mathbf{S}_O$ are the so called spin and orbital flow densities (spin and orbital currents) recently studied in detail [22–24]. By using the Gaussian system of units and denoting the light velocity as $c$ and the wave number as $k$, the summands of Eq. (1) are represented by expressions

$$\mathbf{S}_C = \frac{c}{16\pi k} \mathrm{Im}\left[\nabla \times \left(\mathbf{E}^* \times \mathbf{E}\right)\right], \quad \mathbf{S}_O = \frac{c}{8\pi k} \mathrm{Im}\left[\mathbf{E}^* \cdot (\nabla)\mathbf{E}\right]. \tag{2}$$

Here $\mathbf{E}$ is the complex electric field (the true electric field strength equals to $\mathrm{Re}\left[\mathbf{E}\exp(-i\omega t)\right]$, where the oscillation frequency $\omega = ck$), $\left[\mathbf{E}^* \cdot (\nabla)\mathbf{E}\right]$ is the invariant Berry notation [22] of the vector differential operation that in Cartesian coordinates reads

$$\left[\mathbf{E}^* \cdot (\nabla)\mathbf{E}\right]_j = E_x^* \frac{\partial E_x}{\partial j} + E_y^* \frac{\partial E_y}{\partial j} + E_z^* \frac{\partial E_z}{\partial j}$$

with $j$ standing for $x$, $y$, $z$. In agreement to (1) and (2), the electromagnetic angular momentum of the beam with respect to the certain reference point with radius-vector $\mathbf{R}_0$ can also be represented as a sum of two terms corresponding to summands of (1),

$$\mathcal{L} = \frac{1}{c^2} \mathrm{Im} \int \left[(\mathbf{R} - \mathbf{R}_0) \times \mathbf{S}\right] d^3 R = \mathcal{L}_C + \mathcal{L}_O, \tag{3}$$

which can be reduced to forms

$$\mathcal{L}_C = \frac{1}{8\pi\omega} \mathrm{Im} \int \left(\mathbf{E}^* \times \mathbf{E}\right) d^3 R, \quad \mathcal{L}_O = \frac{1}{8\pi\omega} \mathrm{Im} \int (\mathbf{R} - \mathbf{R}_0) \times \left[\mathbf{E}^* \cdot (\nabla)\mathbf{E}\right] d^3 R. \tag{4}$$

Here $\mathbf{R}$ is the radius vector of the current point of 3D space, the integration is performed over the whole space and it is supposed that $\mathbf{E} \to 0$ rapidly enough at $|\mathbf{R}| \to \infty$.

As is seen from Eqs. (4), term $\mathcal{L}_C$, in contrast to $\mathcal{L}_O$, essentially involves the vector nature of the light wave and does not depend on the reference point position, which properties it shares with the spin AM of a paraxial beam [1]. Moreover, in case of a paraxial beam propagating, say, along axis $z$, the expression of $\mathcal{L}_C$ following from (4) coincides with the usual spin AM definition [1,4,25], so it can be referred to as the "non-paraxial spin AM". The similar but opposite arguments allow the term $\mathcal{L}_O$ to be considered as the orbital AM of a non-paraxial beam. When a paraxial beam is tightly focused, its total AM (3) conserves but the initial well-defined paraxial spin and orbital AM are generally redistributed between the non-paraxial spin and orbital AMs (4) of the focused beam. This effect is commonly treated as the spin-to-orbital AM conversion.

In experiment, the spin and orbital AM, both in the paraxial and non-paraxial versions, can be discriminated by the behavior of absorbing or reflecting particles suspended within the field of the tested light beam. Under the spin AM action, a particle can only rotate near its own axis, regardless of its position within the beam cross section, while in the optical field with orbital AM, particles shifted from the beam axis can perform the orbital motion around it [7–9]. Observation of such orbital motion is just the main experimental evidence that the spin to orbital AM conversion takes place in the strongly focused beams [14,15].

However, this deduction looses sight of the fact that the spin AM per se can also induce the orbital motion of a particle, even in the paraxial case. This conclusion readily follows from the recent analyses of energy flows in light beams [22,23]. In this note, we intend to accentuate this fact and to demonstrate its possible manifestations in usual experimental approaches designed to perform the optically-induced rotations (optical spanners) [7–10].

Let us consider a paraxial light beam propagating along axis $z$. The electric vector distribution of this beam can be represented as [23,24]

$$\mathbf{E} = \mathbf{E}_\perp + \mathbf{e}_z E_z = \exp(ikz)\left(\mathbf{u} + \frac{i}{k}\mathbf{e}_z \operatorname{div}\mathbf{u}\right) \tag{5}$$

where slowly varying vector complex amplitude $\mathbf{u} = \mathbf{u}(x, y, z)$ is related to complex amplitudes of orthogonal polarization components of the field (5), $\mathbf{e}_z$ is the unit vector of longitudinal direction. In the circular polarization basis

$$\mathbf{e}_\sigma = \frac{1}{\sqrt{2}}(\mathbf{e}_x + i\sigma\mathbf{e}_y)$$

($\mathbf{e}_x$, $\mathbf{e}_y$ are unit vectors of the transverse coordinates, $\sigma = \pm 1$ is the photon spin number, or helicity),

$$\mathbf{u} = \mathbf{e}_{+1} u_{+1} + \mathbf{e}_{-1} u_{-1}, \tag{6}$$

$u_\sigma \equiv u_\sigma(x, y, z)$ is the scalar complex amplitude of the corresponding circularly polarized component. Note that in the component with $\sigma = 1$, the electric vector rotates counter-clockwise when seeing against the beam propagation (left polarization in terminology accepted in optics [26]). 'Partial' intensity and phase distributions of each polarization component equal to

$$I_\sigma(x, y, z) = \frac{c}{8\pi}|u_\sigma(x, y, z)|^2, \tag{7}$$

and

$$\varphi_\sigma = \frac{1}{2i}\ln\frac{u_\sigma}{u_\sigma^*}. \tag{8}$$

The spin flow density (2) of the paraxial field (5) reduces to

$$\mathbf{S}_C = \frac{1}{2k}\left[\mathbf{e}_z \times \nabla(I_{-1} - I_{+1})\right] = \frac{1}{2k}\operatorname{rot}\left[\mathbf{e}_z(I_{+1} - I_{-1})\right] = \frac{1}{2k}\operatorname{rot}(\mathbf{e}_z s_3) \tag{9}$$

(see Ref. [23]) where $s_3$ is the fourth Stokes parameter characterizing the degree of circular polarization [27]. Eq. (9) means that, although in transversely uniform beams the circular polarization produces no macroscopic energy current [4,28,29], the specific energy flow occurs in beams with inhomogeneous $s_3$. In particular, this flow is of circulatory character near extrema of function $s_3(x, y)$ [23,24].

The situation becomes especially suitable for analysis in the wide-spread case of a beam with uniform circular polarization and a circular intensity profile. Then $s_3 = \sigma I_\sigma$ and in the polar frame

$$r = \sqrt{x^2 + y^2}, \quad \phi = \arctan(y/x),$$

the corresponding spin flow (9) is expressed by formula

$$\mathbf{S}_C = -\frac{\sigma}{2k}\left(-\mathbf{e}_r \frac{1}{r}\frac{\partial}{\partial \phi} + \mathbf{e}_\phi \frac{\partial}{\partial r}\right) I_\sigma \tag{10}$$

where the unit vectors of polar coordinates are introduced in agreement with equations

$$\mathbf{e}_x = \mathbf{e}_r \cos\phi - \mathbf{e}_\phi \sin\phi, \quad \mathbf{e}_y = \mathbf{e}_r \sin\phi + \mathbf{e}_\phi \cos\phi.$$

For comparison, the orbital flow density (transverse part of the second expression (2)) of the same beam, in accord with (5) – (8), is given by equation [23]

$$\mathbf{S}_O = \frac{1}{k} I_\sigma \nabla \varphi_\sigma = \frac{1}{k} I_\sigma \left(\mathbf{e}_\phi \frac{1}{r}\frac{\partial}{\partial \phi} + \mathbf{e}_r \frac{\partial}{\partial r}\right)\varphi_\sigma. \tag{11}$$

For paraxial beams, it is natural to consider the AM with respect to the propagation axis $z$ and to characterize it by the linear density (AM per unit length of the beam) [1,4,11] which is expressed by the proper modification of Eq. (3) [4,23]

$$\mathcal{L}' = \frac{1}{c^2} \mathrm{Im} \int [\mathbf{r} \times \mathbf{S}] d^2 r = \frac{1}{c^2} \mathrm{Im} \int S_\phi r^2 \, dr d\phi$$

where $\mathbf{r}$ is the transverse radius-vector, $S_\phi$ is the Poynting vector azimuthal component and the integration is performed over the whole cross section of the beam. With allowance for Eqs. (10) and (11), the spin and orbital AM linear densities for a paraxial beam can be written in the well known forms (see, e.g., [23])

$$\mathcal{L}'_C = -\frac{\sigma}{2\omega c} \int_0^\infty r^2 dr \int_0^{2\pi} \frac{\partial I_\sigma}{\partial r} d\phi = \frac{\sigma}{\omega c} \int_0^\infty r dr \int_0^{2\pi} I_\sigma d\phi, \quad (12)$$

$$\mathcal{L}'_O = \frac{1}{\omega c} \int_0^\infty r dr \int_0^{2\pi} I_\sigma \frac{\partial \varphi_\sigma}{\partial \phi} d\phi \quad (13)$$

(in the second Eq. (12) the fact that $I_\sigma(r,\phi) \to 0$ when $r \to \infty$ has been employed).

One can remark a great degree of similarity between Eqs. (10) and (11): both $\mathbf{S}_C$ and $\mathbf{S}_O$ originate from the beam transverse inhomogeneity and their components are directly related to the azimuthal and radial derivatives of the beam profile parameters. However, while the orbital flow is mainly "produced" by the phase gradient (and the variable intensity can only modify it due to factor $I_\sigma$), the spin flow completely owes to the amplitude inhomogeneity of a circularly polarized beam. There also exists a difference in the interrelations between the stream line patterns of $\mathbf{S}_C$ ($\mathbf{S}_O$) and the spatial derivatives of the corresponding "master" parameter $I_\sigma$ ($\varphi_\sigma$): while $\mathbf{S}_O$ is always directed *along* the *phase* gradient, $\mathbf{S}_C$ is *orthogonal* to the *intensity* gradient. Nevertheless, in what concerns the action on suspended microparticles, both flows are expected to be quite equivalent provided that the quantitative characteristics of the flow patterns are commensurable. Now consider the detailed characterization of the spin and orbital flows in some simple examples.

For a Gaussian beam in the waist cross section (beam waist radius $b$) the intensity (7) and phase (8) distributions appear in the forms

$$\varphi_\sigma = 0, \; I_\sigma = I_{\sigma 0} \exp\left(-\frac{r^2}{b^2}\right). \quad (14)$$

The wavefront of this beam is flat and the orbital flow (11) vanishes; the spin flow is determined by the last term of (10)

$$\mathbf{S}_C = -\sigma \mathbf{e}_\phi \frac{1}{2k} \frac{\partial I_\sigma}{\partial r} \quad (15)$$

which due to Eq. (14) gives

$$\mathbf{S}_C = \sigma \mathbf{e}_\phi \frac{r}{kb^2} I_{\sigma 0} \exp\left(-\frac{r^2}{b^2}\right) \quad (16)$$

(in Fig. 1 the spin flow pattern in the left-polarized beam, $\sigma = 1$, is presented).

A bit more complicated situation occurs in Laguerre-Gaussian (LG) beams which along with the spin helicity (circular polarization) possess the "orbital helicity" – the screw wavefront dislocations giving rise to optical vortices of $l$-th order ($|l| > 1$ is the integer azimuthal index) [1,4,11]. Being restricted, for simplicity, by beams with zero radial index, let us again consider the waist cross section where

$$\varphi = l\phi, \; I_\sigma = \frac{1}{|l|!} I_{\sigma 0} \left(\frac{r}{b}\right)^{2|l|} \exp\left(-\frac{r^2}{b^2}\right). \quad (17)$$

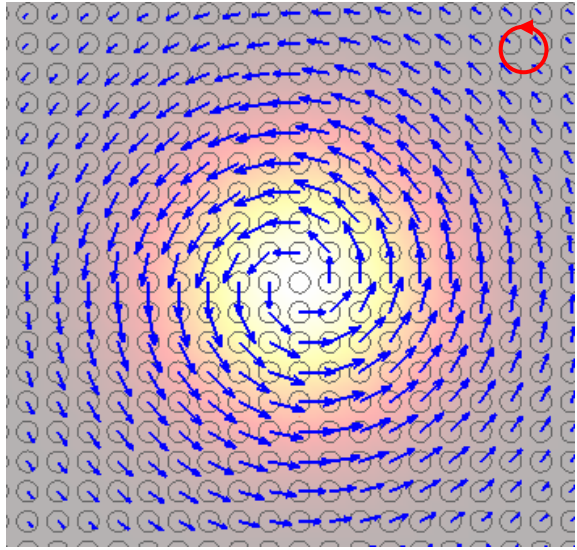

Fig. 1. Map of the spin flow density of Eq. (16) for a left-polarized Gaussian beam ($\sigma = 1$, polarization handedness is shown in the upper right corner); lengths of arrows correspond to relative flow density, the intensity distribution and polarization ellipses (circles) are shown in the background, the beam is viewed against the propagation axis.

The normalization constant $(|l|!)^{-1}$ warrants that the beam total power for every $l$ is the same. With allowance for (17) the last term of (10) gives

$$\mathbf{S}_C = -\mathbf{e}_\phi \sigma I_{\sigma 0} \frac{1}{|l|!} \frac{1}{kb} \left(\frac{r}{b}\right)^{2|l|-1} \left(|l| - \frac{r^2}{b^2}\right) \exp\left(-\frac{r^2}{b^2}\right), \tag{18}$$

and, following to (11), the orbital flow is found to be

$$\mathbf{S}_O = \mathbf{e}_\phi I_{\sigma 0} \frac{1}{|l|!} \frac{1}{kb} \left(\frac{r}{b}\right)^{2|l|-1} l \exp\left(-\frac{r^2}{b^2}\right). \tag{19}$$

These equations stipulate a simple relation between the spin and orbital flows of the considered circularly polarized beams:

$$\mathbf{S}_C = -\frac{\sigma}{l}\left(|l| - \frac{r^2}{b^2}\right)\mathbf{S}_O \quad (l \neq 0). \tag{20}$$

The derived dependencies are illustrated by Figs. 2a–d. In contrast to the spin and orbital AM densities (12), (13) which usually coincide [1,4] with the transverse intensity distribution of circularly polarized LG beams (curves $I$), the corresponding transverse energy flows (curves $S_C$ and $S_O$) behave differently. At any $l$, the circulatory energy flows vanish on the axis ($r = 0$); of course, far from the axis ($r \to \infty$) they vanish as well. In the intermediate region absolute values of the spin and orbital flows possesses extrema. The orbital flow magnitude (19) has the maximum at

$$\frac{r}{b} = \sqrt{\frac{2|l|-1}{2}} \quad (|l| > 0), \tag{21}$$

extremum points of the spin flow density (18) satisfy the condition

$$\left(\frac{r}{b}\right)^2 = |l| + \frac{1}{4} \pm \frac{\sqrt{16|l|+1}}{4} \tag{22}$$

which corresponds to maximums of $|\partial I_\sigma / \partial r|$ on the inner and outer sides of the bright ring of the "doughnut" mode pattern (17); at $l = 0$ the inner extremum disappears and the only maximum of the absolute spin flow density occurs at

$$r = b/\sqrt{2}.\tag{23}$$

The expectable zero spin flow takes place at the "brightest" line of the ring where $I_\sigma$ is maximal.

Eqs. (18), (19) and Fig. 2 show that in many cases magnitudes of the spin and orbital flow densities are of the same order. Consequently, they are expected to have similar experimental manifestations. In particular, since the orbital flow, due to associated mechanical momentum, can force the orbital rotation of particles [7,8], the same effect can be caused by the spin flow. This must be taken into account in experiments on the spin-to-orbital AM conversion [14,15]. In real situations, it is the total transverse energy flow

$$\mathbf{S} = \mathbf{S}_C + \mathbf{S}_O = \left(1 - \sigma \frac{|l|}{l} + \frac{\sigma}{l}\frac{r^2}{b^2}\right)\mathbf{S}_O \tag{24}$$

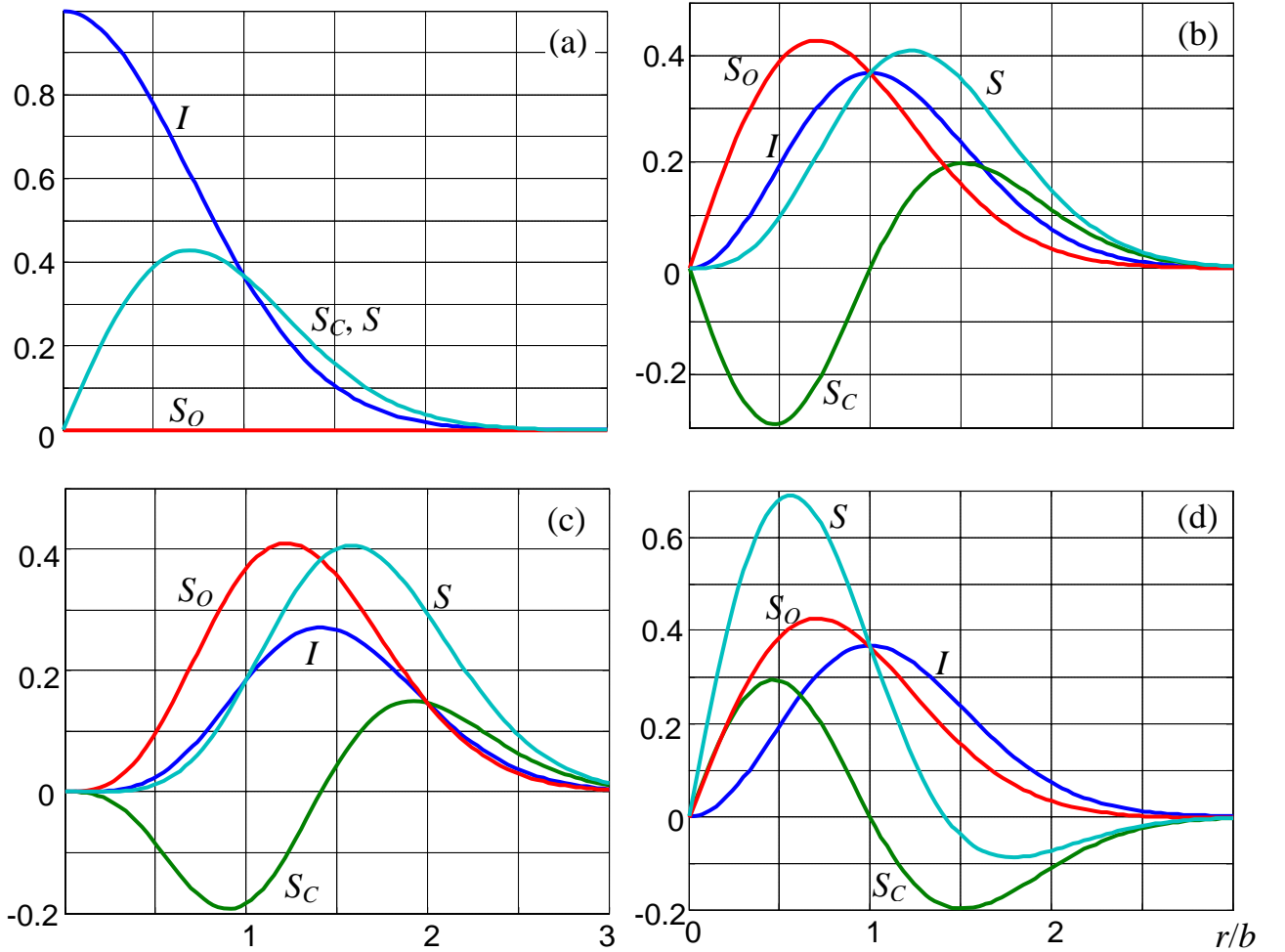

Fig. 2. Radial profiles of ($I$) intensity (17) in units of $I_{\sigma 0}$, ($S_C$) spin flow density (18), ($S_O$) orbital flow density (19) and ($S$) total transverse flow density (24) (all in units of $I_{\sigma 0}/kb$), for the circularly polarized LG beams with zero radial index and the following sets of parameters: (a) $\sigma = 1$, $l = 0$ (Gaussian beam of Fig. 1), (b) $\sigma = 1$, $l = 1$, (c) $\sigma = 1$, $l = 2$, (d) $\sigma = -1$, $l = 1$.

(Fig. 2, curves marked $S$), with associated mechanical momentum $\mathbf{P} = \mathbf{S}/c^2$, that is likely to be the motive factor for orbital rotation of the probing particles. The spin and orbital contributions may support as well as suppress each other (see Fig. 2). In the region $r/b < l$, the most important physically because it contains prevailing part of the beam power, the orbital flow dominates; otherwise (at the beam periphery) the spin contribution is more intensive.

An interesting situation occurs in the near-axis region $r/b \ll 1$ where, due to Eq. (20), absolute magnitudes of the spin and orbital flows are almost identical. Then, if signs of $l$ and $\sigma$ coincide (that is, handedness of the macroscopic optical vortex of the LG beam and handedness of the circular polarization are the same), the total transverse energy circulation is zero at small $r \ll b$ (see Fig. 2b). That the spin flow can be directed oppositely to the polarization handedness, seems, at first sight, counter-intuitive but can be simply explained by the "cell model" of the spin flow formation [4,29]. Formally, this follows immediately from the fact that the spin flow handedness is determined not only by $\sigma$ but also by the sign of $\partial I_\sigma / \partial r$ (15).

On the contrary, if the polarization handedness is opposite to the orbital circulation, the spin and orbital flows add constructively and enable the maximum local values of the total rotational energy flow available for circularly polarized LG beams with given $l$, as is seen from Fig. 2d, curve $S$.

The flow maps presented in Fig. 3 are in full agreement with the data of Fig. 2d. For considered beams, the orbital flow density possesses the same handedness in the whole cross section (compare Fig. 3a and curve $S_O$); however, the spin and the total flows may reverse. Regions of opposite circulations in Figs. 3b, c are separated by contours where the relevant energy flow constituent vanishes, which correspond to sign alterations in curves $S_C$ and $S$.

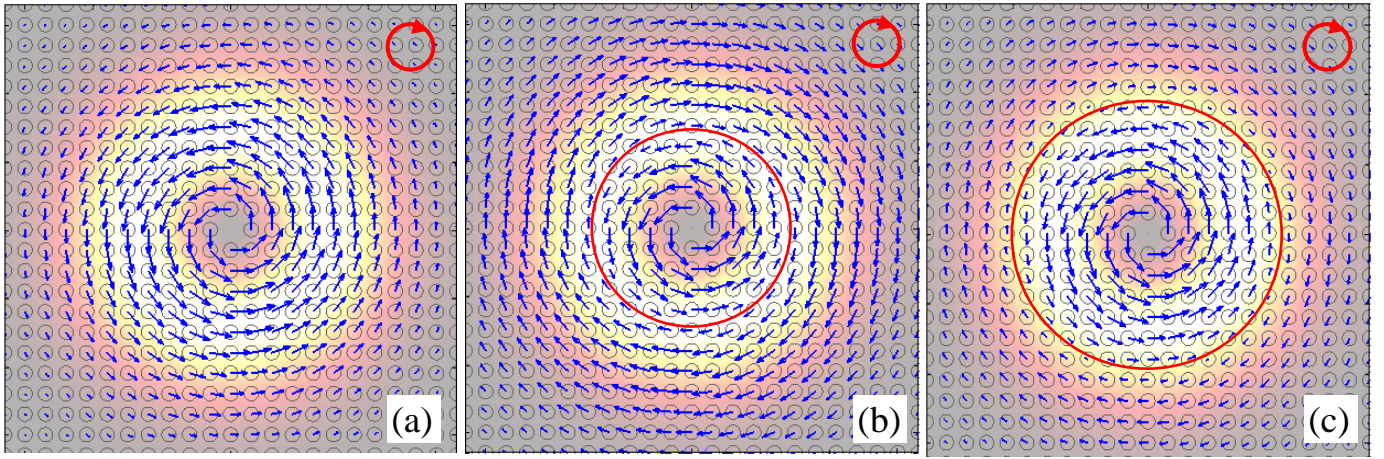

Fig. 3. Maps of the (a) orbital $\mathbf{S}_O$ (b) spin $\mathbf{S}_C$ and (c) total $\mathbf{S}$ transverse energy flows in the cross section of a right-polarized LG beam (17) with $l = 1$, $\sigma = -1$ (case of Fig. 2d). In every point, polarization is the same as shown in the upper right corners; circular contours in panels (b) and (c) are contours where the corresponding flow component vanishes.

Note that in calculation of the full spin AM over the whole cross section (e.g., by first formula (12)), the "opposite" spin flow of the near-axis region is compensated by the periphery contribution where the spin flow reverses. As a result, the handedness of the total spin AM of the considered uniformly polarized beam always coincides with $\sigma$, which is seen from the second Eq. (12) where this compensation is ensured automatically.

Now let us dwell upon the peculiarities of the spin flow as a factor inducing the orbital motion of suspended particles and experimental conditions enabling unambiguous manifestation of the spin flow. Its action would be especially expressive in case of Gaussian beam (14) where the orbital flow is absent. In the most general features, the idea of experiment does not differ much from that devised for the orbital AM demonstration [9,10,14]. The tested beam falls normally onto the cell with suspended particles which are situated off-axially with respect to the beam axis. They experience the motive force proportional to the local energy flow density expressed, for example, by Eqs. (15), (16), (18) and (19). In fact, this force is directed tangentially and, if it is the only force acting on the particles, they move centrifugally. In order to get the closed orbital motion, some additional steps should be taken to keep the particles at a fixed circular trajectory. In experiments with beams carrying the orbital AM this can be realized due to special configuration of the beam itself. If it possesses a (multi-)ring-like transverse profile (e.g., LG beam with at least one non-zero index, Bessel beam, etc.), the particles undergo the gradient force due to the optical field inhomogeneity, and tend to be confined within rings of high or low intensity depending of their optical properties. In studies of the spin flow this technique is inappropriate because in regions of the intensity extrema the spin flow vanishes (see Eq. (15)). This forces to look for other solutions (see Fig. 4). For example, the cell with suspended particles may contain a ring-like channel or cuvette of the proper mean radius corresponding to the maximum spin flow (Fig. 4a, b), e.g., what is dictated by Eq. (22) or (23). In the channel, the particles are kept mechanically, e.g. due to special shape of the cell bottom. Such a mechanical trapping may be inconvenient because the particles' orbital motion is hampered by friction at the channel boundaries. Otherwise, the channel can be formed by a sort of ring-like optical trap, for example, by an auxiliary light beam with ring-like intensity profile. The intensity of the auxiliary field must be sufficient to form the perceptible peak or gap in the resulting intensity distribution (with account for the driving beam whose spin flow is analyzed); besides, the auxiliary beam should be free from additional rotatory action (possess no orbital AM).

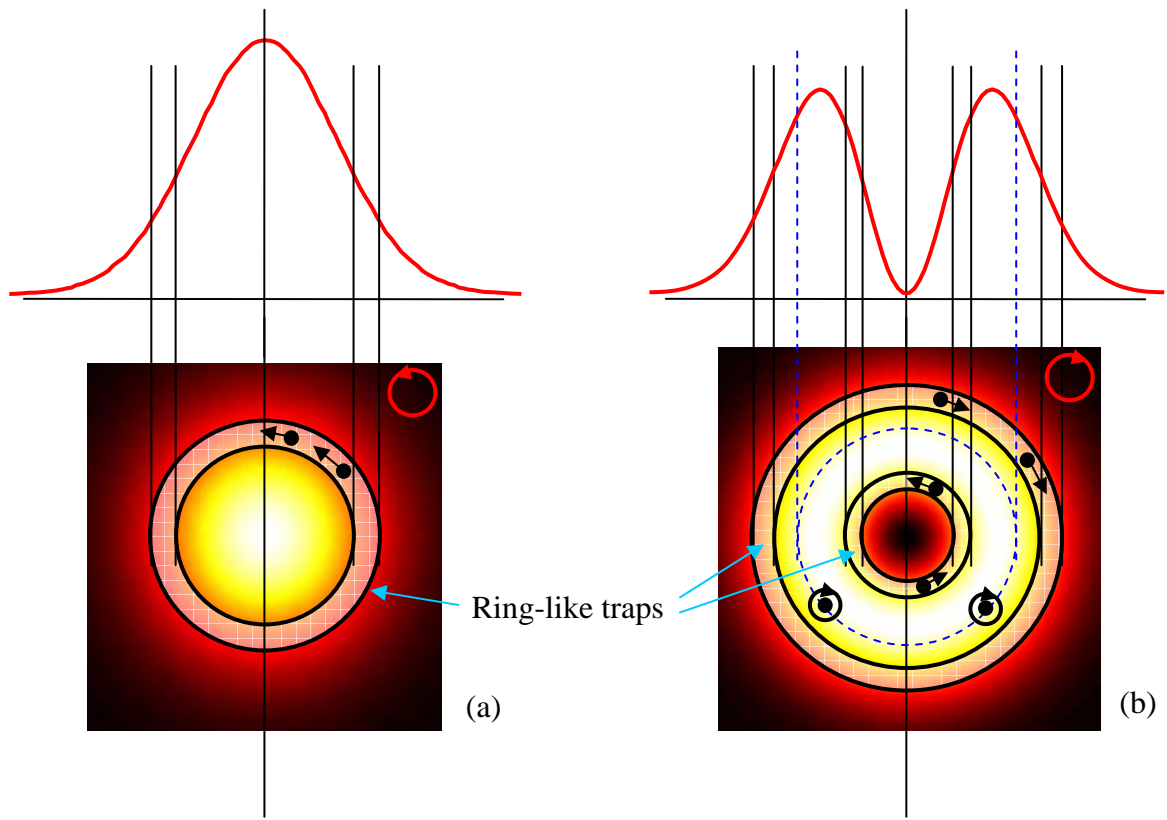

Fig. 4. Possible schemes of orbital motion of the absorbing suspended particles confined in the ring-like traps within the circularly polarized field of (a) Gaussian beam of Fig. 1 and (b) LG beam with $l = 1$, $\sigma = -1$ of Fig. 3. Top row: diametral sections of the intensity profiles with boundaries of the ring-like traps, bottom row: views of the beam cross sections with the traps' traces (polarization handedness is indicated in the upper right corners). Circles with arrows indicate the expected orbital motion of the trapped particles, dashed lines in panel (b) specify locations where orbital motion is not excited (cf. the circular contour in Fig. 3c). Particles situated at this contour perform only the spinning motion (shown by the arrow loops); in all other positions the spinning motion is not shown but is also expected in addition to the orbital one.

Interesting possibilities open up due to variable handedness of the transverse energy circulation, as Fig. 3c displays. This pattern means that direction of the tangential force applied to a particle depends on its radial position so the speed and direction of the orbital rotation can be switched by changing the driving beam radius or the ring-like trap radius. Another expected peculiarity of the motion caused by the spin flow is that the particles absorbing a part of the incident circularly polarized light will thus be set in rotation about their own axes, in addition to the orbital motion around the driving beam axis (see Fig. 4b). Handedness of this spinning motion is the same over the whole cross section of the homogeneously polarized beam, although its rate will generally vary in accord with the inhomogeneous intensity.

The relatively simple examples considered in this work illustrate the main properties and potentiality of the spin flow in problems of optical manipulation. Even if not employed, the spin constituent of the transverse energy circulation must be taken into account in experiments involving the optically induced orbital rotation of microparticles. Interesting applications of circularly polarized beams may arise from the possibility of combining the orbital and spinning motion of the same particle. A more reach variety of available particle's motions and new possibilities of their control can be expected in case of more complicated driving beams, e.g. those with inhomogeneous polarization. The presented analysis constitutes the starting point and outlines the way in which these complicated cases can be studied further.